\documentclass[letterpaper, 10 pt, conference]{ieeeconf}  

\IEEEoverridecommandlockouts                              

\usepackage{mathptmx}       
\usepackage{helvet}         
\usepackage{courier}        
\usepackage{type1cm}        
\usepackage{makeidx}         
\usepackage{graphicx}        
\usepackage{multicol}        
\usepackage[bottom]{footmisc}
\usepackage{amsmath} 
\usepackage{mathrsfs}
\usepackage{amssymb}
\usepackage{graphicx} 
\usepackage{epstopdf}
\usepackage{wrapfig}
\usepackage{color}
\usepackage{gensymb}
\usepackage{bm}
\usepackage{mathtools}
\graphicspath{ {Figures/} }
\usepackage{epstopdf}
\usepackage[mathlines]{lineno}
\usepackage{caption}
\usepackage{setspace}
\usepackage{float}
\usepackage{dcolumn}
\usepackage{bm}
\usepackage{url}

\def\be{\begin{eqnarray}}
\def\ee{\end{eqnarray}}
\def\bes{\begin{subeqnarray}}
	\def\ees{\end{subeqnarray}}

\def\lp{\left(}
\def\rp{\right)}

\def\ba#1\ea{\begin{align}#1\end{align}}

\usepackage{scrextend}
\deffootnote[0em]{1.5em}{1.5em}{\textsuperscript{\thefootnotemark}\,}

\title{\LARGE \bf
The Experimental Realization of an Artificial Low-Reynolds-Number Swimmer with Three-Dimensional Maneuverability$^*$}

\author{Mohsen Saadat$^{1}$, Mehdi Mirzakhanloo$^{1}$, Julie Shen$^{1}$, Masayoshi Tomizuka$^{1}$, and Mohammad-Reza Alam$^{1}$ 
\thanks{*This work is supported by the National Science Foundation under grant CMMI-1562871.}
\thanks{$^{1}$ The authors are with the Department of Mechanical Engineering at the University of California, Berkeley, CA 94720, USA. Emails:\{\tt\small  {msaadat, m.mirzakhanloo, jshen98,  tomizuka, reza.alam\}@berkeley.edu}}%
}

\begin{document}

\maketitle
\thispagestyle{empty}
\pagestyle{empty}

\begin{abstract}

The motion of biological micro-robots -- similar to that of swimming microorganisms such as bacteria or spermatozoa -- is governed by different physical rules than what we experience in our daily life. This is particularly due to the low-Reynolds-number condition of swimmers in micron scales. The Quadroar swimmer, with three-dimensional maneuverability, has been introduced for moving in these extreme cases: either as a bio-medical micro-robot swimming in biological fluids or a mm-scale robot performing inspection missions in highly viscous fluid reservoirs. Our previous studies address the theoretical modeling of this type of swimmer system. In this work, we present the mechatronic design, fabrication, and experimental study of a mm-scale Quadroar swimmer. We describe the design methodology and component selection of the system based on the required performance. A supervisory control scheme is presented to achieve an accurate trajectory tracking for all the actuators used in the swimmer. Finally, we have conducted experiments in silicone oil (with 5000 cP viscosity) where two primary modes of swimming - forward translation and planar reorientation - have been tested and compared with the theoretical model.

\end{abstract}

\section{INTRODUCTION}

Although the physical laws of motion are universally governed by Newton's equation, environment plays a crucial role in developing the locomotory functions of animals and organisms. Locomotion is a result of gaining reaction forces from an environmental entity by the continuous movements of body parts. In walking or running, a reaction force is generated at the contact point with ground by pushing the Earth backwards, while in swimming and flying, animals self-propel by displacing their environmental fluid. 
Although the flying of birds and swimming of fish involve periodic movements of their body parts, displacing the environmental fluid becomes harder when viscous forces overwhelm inertial ones. Specifically, in the limit of zero Reynolds number, when the motion of the background fluid is governed by the Stokes equation, periodic movements of the body parts will not accelerate the organism \cite{Purcell77,Happel12}. The same problem exists for robotic swimmers in these extreme conditions. 

Several designs can be found for artificial low-Reynolds-number swimmers in the literature: multiple-link mechanisms \cite{Purcell77}, linked spheres \cite{Najafi04}, swimmers with helical propellers \cite{Zhang09}, and the Quadroar \cite{Jalali2014}. Among these systems, linked spheres are hard to manufacture (especially three-dimensional ones) because they need assemblies of linear actuators with relatively long stroke lengths. Helical swimmers are the easiest to realize using 3D printers that use two-photon polymerization \cite{Tottori12}, but they have only one degree of freedom and path planning (as one may need to realize concealed swarms \cite{Mirzakhanloo2019}) for them is impossible. Multiple-linked mechanisms are easier to manufacture and operate because they need only rotary actuators, but their motion is confined to 2D space -- to the best of our knowledge, swimmers with 3D linkages have not been introduced. 

\begin{figure}[h]
	\centering
	\includegraphics[width=0.38\textwidth]{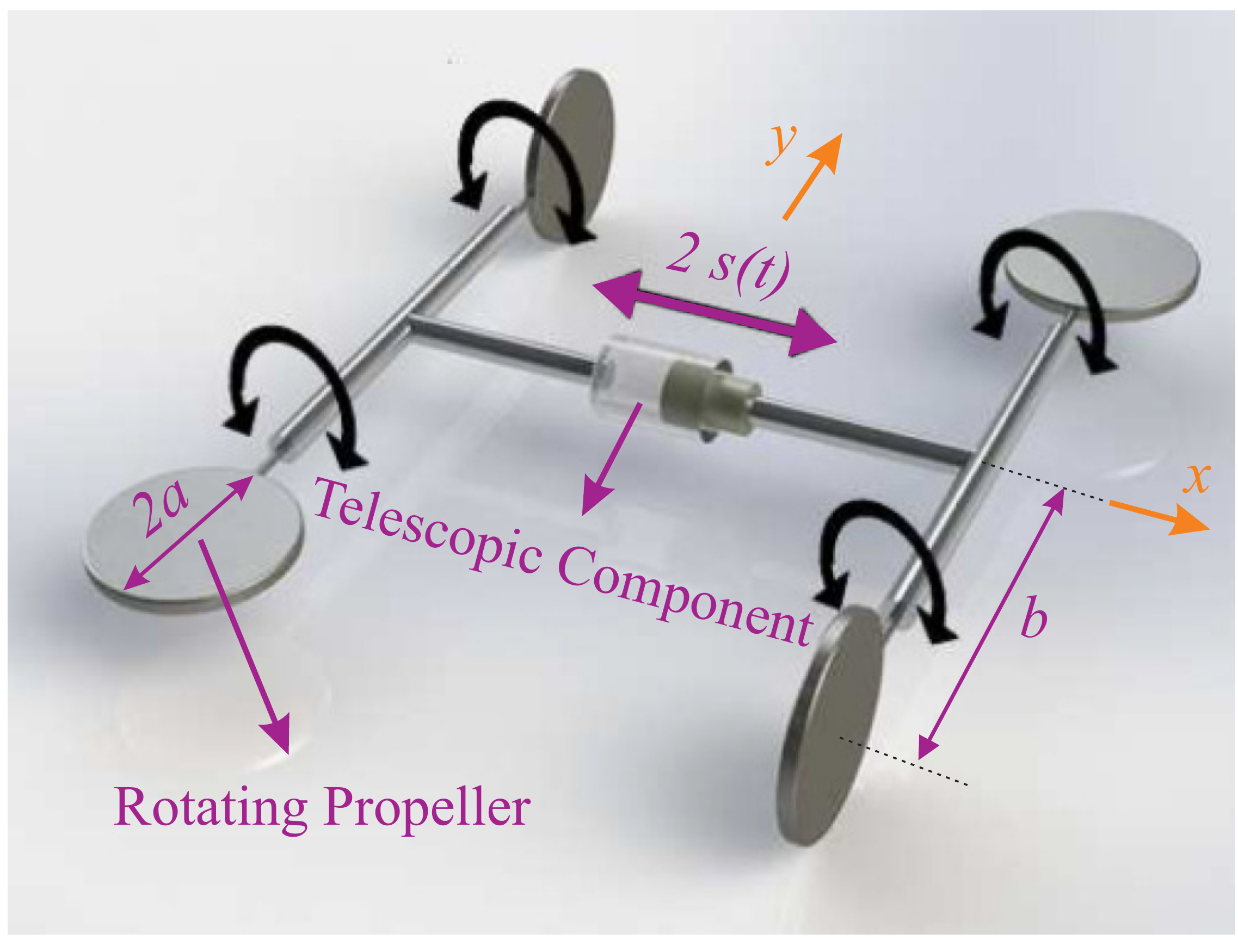} 
	\caption{Schematic representation of the Quadroar swimmer \cite{Jalali2014}, which consists of four rotating disks, and a linear actuator along its chassis.
	}
	\label{fig1}
\end{figure}

The Quadroar swimmer has four rotating paddles and a chassis that can expand and contract linearly (see Fig. \ref{fig1}). 
Possible pathways towards the realization of the Quadroar in micron and sub-micron scales have been discussed in Ref. \cite{Jalali2014}. In this paper, we report the design, fabrication and control of a mm-scale Quadroar. The swimmer's paddles and linear actuator are equipped with feedback control systems and Hall effect sensors. It can transmit data with a central computer through a USB cable. As the final step, we have tested the swimmer prototype in silicone oil (to achieve a low Reynolds number regime) and studied its behaviour for two primary modes of operation: (i) Forward Translation; and (ii) Planar Reorientation (see Fig. \ref{fig1_2}).

Potential applications of this artificial low-Reynolds swimmer range from robotic inspection of oil tanks and environmental monitoring/remediation, to multi-agent drug delivery systems through blood vessels. Our artificial micro-swimmer can induce chaotic mixing \cite{Jalali2015} that may be used to revitalize marine life in dead zones of the oceans and estuaries where the shortage of microorganisms and their stirring contribution (c.f. \cite{Katija2012,Wilhelmus2014}) have negatively impacted the ecological cycle. A flock of micro-Quadroars injected into the circulatory system may prevent clot formation in patients susceptible to thrombosis. It may also speed up drug absorption rate and efficiency that is critical under emergency circumstances. 

This manuscript is organized as follows. Section 2 presents theoretical background on the Quadroar swimmer. The mechatronic design and fabrication of a mm-scale swimmer based on the Quadroar architecture is described in section 3. Controller design methodology is introduced in section 4. Section 5 presents the experimental results achieved by testing the swimmer system in a pool of silicone oil. Concluding remarks are given in section 6.


\section{THEORETICAL BACKGROUND}
\label{theory}

Schematic of the Quadroar (\textit{Quadru$+$oar}) swimmer is shown in Fig.\ref{fig1}. It consists of an I-shaped frame, including two axles and a middle telescopic link. Disks of radii $a$ are placed at the ends of each axle which can rotate around their supporting flange link. The orientation of the i$^{th}$ (i$\in \{1,2,3,4\}$) disk with respect to the main frame (or swimmer chassis) is denoted by the angle variable $\theta_i$, and the length of each axle is described by $2b$. The telescopic link has a variable length $2l+2s\lp t \rp$, where $s\lp t \rp$ is the contribution of the expansion-contraction of the linear actuator. Orientation of the swimmer is measured by the roll-pitch-yaw sequence of Euler's angles ($\bm{\alpha}$), and its motion is tracked by the position of its geometric center $\bm{X}_c$. Local frame of reference is fixed to the geometric center of the swimmer, so that the I-frame lies in the $x-y$ plane, and $x$-axis is along the chassis.

As a micro-robot swimming in biological fluids or a mm-scale robot performing inspection missions in highly viscous fluid reservoirs, the Quadroar swimmer is a low-Reynolds-number swimmer. In low-Reynolds regime, inertia is negligible and viscous drags primarily govern the dynamics of the system. In an otherwise stationary fluid of viscosity $\mu$, the i$^{th}$ paddle (i.e., disk) of the swimmer experiences a drag force and drag torque, while translating and rotating. These forces and torques are functions of the fluid viscosity ($\mu$), disk geometry, and its linear and angular velocities (denoted by $\bm{v}$ and $\bm{\omega}$). Mathematically:
\ba \label{eq1}
\bm{f}_{i} = \mu \ \mathscr{K}_{i} \cdot \bm{v}_{i}, \quad  \bm{\tau}_{i} =\mu \ \mathscr{G} \cdot \bm{\omega}_{i} 
\ea
Where all geometric properties of the disk are gathered in two tensors: the translation tensor $\mathscr{K}_{i}$ and the isotropic tensor $\mathscr{G}$ \cite{Jalali2014}. A self-propelled swimmer in low-Reynolds regime must be force-free and torque-free. Thus, balance of force and torque require $\mu \sum_{i=1}^{4} \mathscr{K}_{i} \cdot \bm{v}_{i} = 0$ and $\mu \sum_{i=1}^{4} \lp G \cdot \bm{\omega}_{i} + \bm{r}_i \times \mathscr{K}_{i} \cdot \bm{v}_{i} \rp  = 0$, where $\bm{r}_i$ is the position vector of disk $i$. Note that absolute linear and angular velocities of each disk (i.e. $\bm{v}_{i},\bm{\omega}_{i}$) can be calculated knowing velocity of the swimmer's center of mass, $\bm{V}_c$, and its angular velocity, $\bm{\Omega}$. Therefore, the force- and torque-balance equations can be combined and expressed as \cite{Jalali2014}:
\be \label{eq2}
\begin{bmatrix} C_{11} & C_{12} \\  C_{21} & C_{22} \end{bmatrix}  \begin{pmatrix} \bm{V}_c \\ \bm{\Omega} \end{pmatrix} = \ \begin{pmatrix} f_1 \\ f_2 \end{pmatrix}.
\ee
Detailed expressions for elements of the resistance matrix, $C_{mn}$, and the forcing vector $f_m$ (m,n $\in \{ 1,2 \}$), are given in our earlier publication \cite{Jalali2014}. Eq. \eqref{eq2} consists of two non-linear ordinary differential equations, with parametric and external excitation. Once this non-linear system is solved for $\bm{V}_c$ and $\bm{\Omega}$, trajectory of the swimmer and its orientation ($\bm{\alpha}$) in time can be found through integrating:
\begin{equation} \label{eq3}
\dot{\bm{X}}_c= \bm{V}_c, \quad \dot{\bm{\alpha}}= T^{-1} \cdot \bm{\Omega},
\end{equation}
where $T$ is the transformation matrix \cite{Jalali2014}. For a detailed description of governing equations, numerical techniques which have been used in order to get the theoretical results, and different operation modes, the reader is referred to our previous studies (see e.g. \cite{Jalali2014,Jalali2015,Mirzakhanloo2018A,Mirzakhanloo2018B}).

In practice, behavior of the swimmer is controlled by five control inputs: rotation rate of the four paddles and motion of the linear actuator. Through these five degrees of freedom, the Quadroar swimmer is able to move along transverse or forward straight lines, and perform re-orientation maneuvers about all of its body-fixed local axes \cite{Jalali2014}. Thus, the swimmer has full 3D maneuverability and is able to track any prescribed path in space by means of step-wise control strategies. This may potentially address some of the main challenges that biological micro-robots are facing today \cite{Medina2017}.

\begin{figure}[h]
	\centering
	\includegraphics[width=0.48\textwidth]{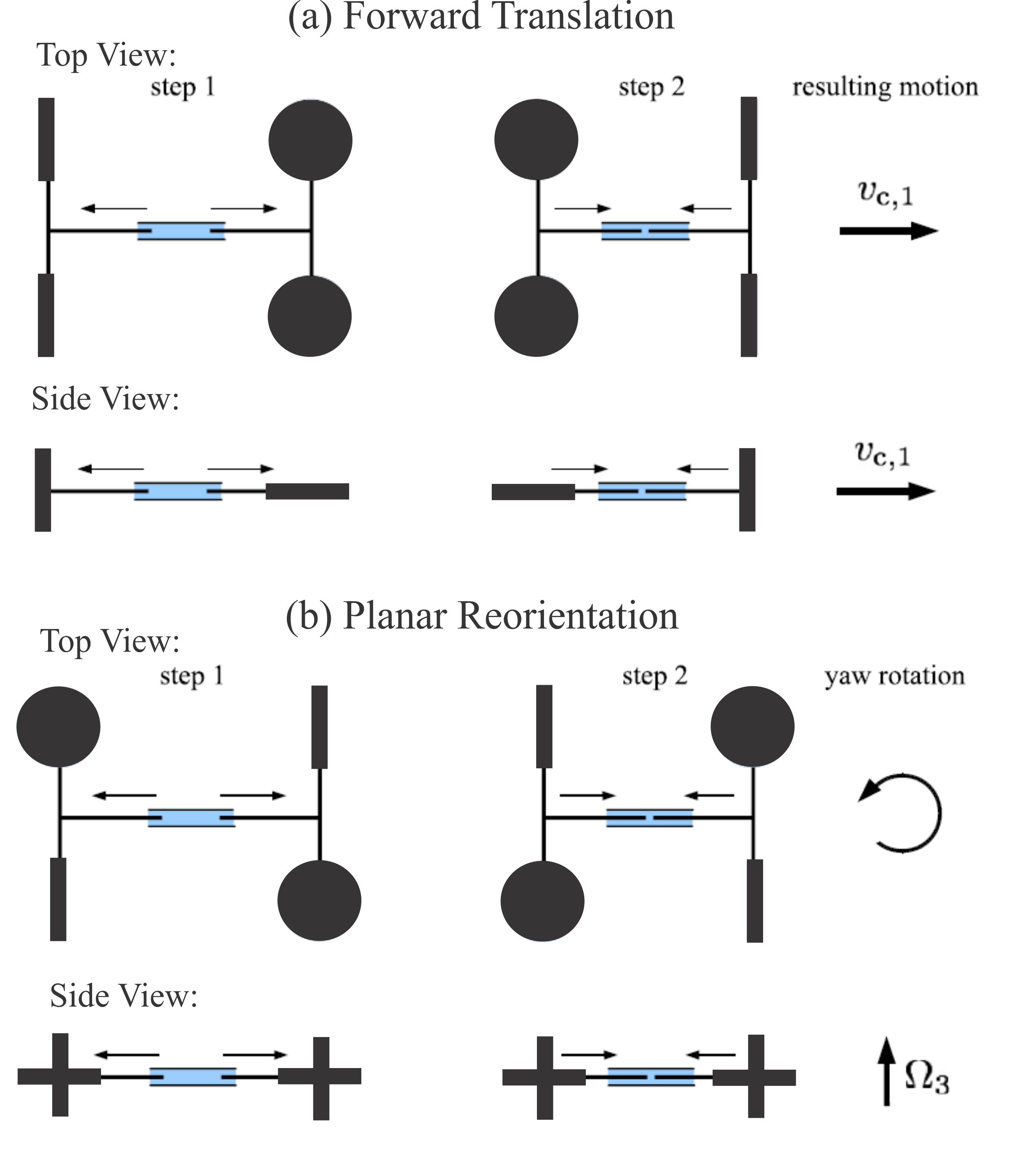} 
	\caption{Schematic representation of stroke cycles leading to forward translation (a) and planar reorientation (b) of the Quadroar swimmer \cite{Jalali2014}.}
	\label{fig1_2}
\end{figure}
Here we consider a planar space, where translating forward along the chassis of the swimmer, combined with the ability to reorient toward any desired direction, makes the swimmer able to track any prescribed path in 2D space. The sequences of step-wise control inputs  (so-called `swimming cycles') which result in these two primary modes of planar motion are presented in Fig. \ref{fig1_2}. Specifically, harmonic expansion-contraction of the linear actuator while keeping the rear disks normal (parallel) to the I-frame, and the front disks parallel (normal) to the I-frame, during the expansion (contraction) results in the `Forward Translation' mode. A similar cycle, except with disks 1\&4 normal (parallel) to, and disks 2\&3 parallel (normal) to the I-frame during expansion (contraction), leads to in-place counter-clockwise reorientation of the swimmer, called `Planar Reorientation' mode.

\section{Design of a \lowercase{mm}-Scale Quadroar Swimmer}
\label{design}

\begin{figure}[h]
	\centering
	\includegraphics[width=0.45\textwidth]{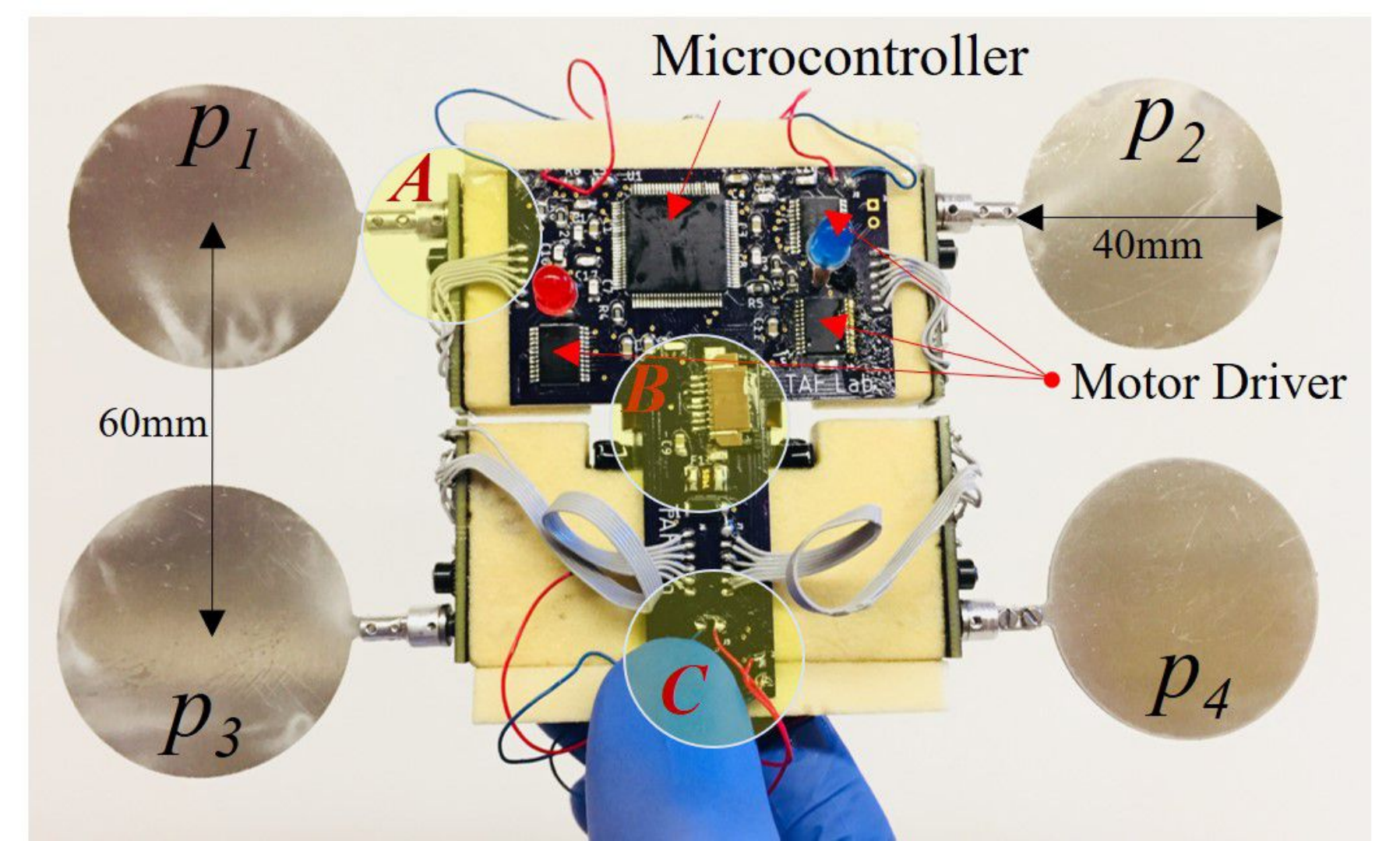} 
	\caption{A mm-scale of the Quadroar swimmer system for working in low-Reynolds number condition. A) Angle measurement system; B) Linear expansion/contraction mechanism; C) Expansion/contraction measurement system.}
	\label{fig0}
\end{figure}

In this work, we designed, fabricated and tested a mm-scale swimmer system according to the proposed Quadroar architecture. We used highly viscous silicone oil to generate a low-Reynolds-number condition since the swimmer size is in mm range. The swimmer system is composed of four rotating disks attached at the ends of its front and rear axles. These axles are located on a variable length chassis (see Fig. \ref{fig0}). Five mm-size brushed DC geared motors are used to spin the disks relative to the swimmer body and to achieve linear expansion/contraction between the front and rear parts of the chassis. Additionally, a sensory system is used to measure the paddle angles and the chassis displacement to implement various modes of operation. An on-board micro-controller is utilized to obtain trajectory tracking for all five actuators. In what follows, we briefly present the design and fabrication of the main mechanical and electrical components used in the Quadroar swimmer system.

\subsection{Main Body and Chassis}
\begin{figure}[h]
	\centering
	\vspace{2mm}
	\includegraphics[width=0.45\textwidth]{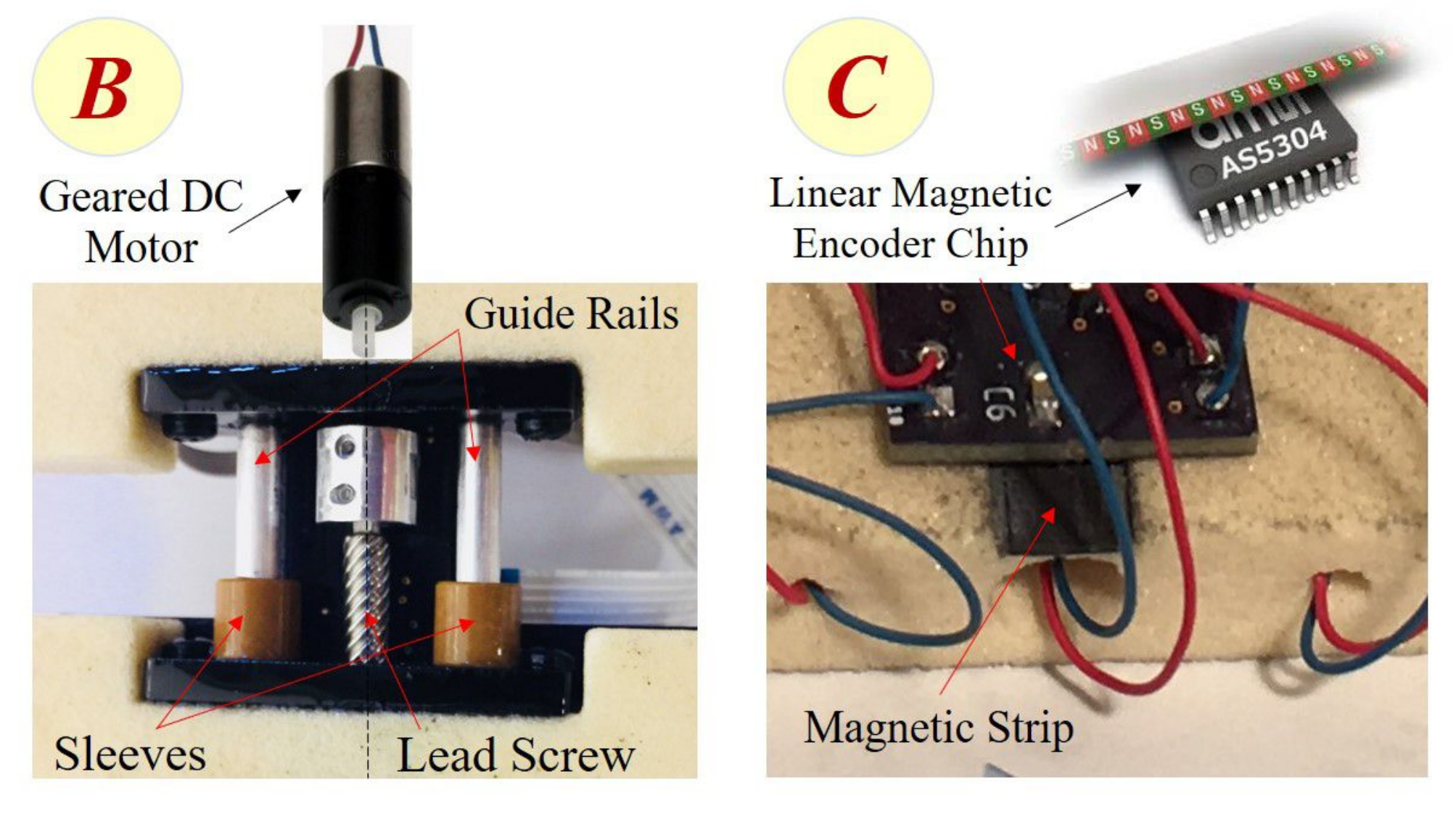} 
	\caption{A lead-screw mechanism is used to generate the expansion/contraction motion between the front and rear sections of the swimmer.}
	\label{fig44}
\end{figure}

The main body of the swimmer consists of two rectangular-shape sections (front and rear). Since most of the components installed on the swimmer (i.e. DC motors, magnetic encoders, lead-screw mechanism, etc.) have a density higher than that of silicone oil, a low-density machinable foam with a density of about 0.2gr/cm$^3$ is used to fabricate the body of the swimmer. The density of the swimmer is approximately 0.95gr/cm$^3$, suitable for submerging in the silicone oil. A lead-screw mechanism with two guide rails and a small geared DC motor to turn the lead-screw is deployed to achieve linear expansion/contraction between the front and rear sections (see Fig. \ref{fig44}).

\subsection{Motors and Motor Drivers}
In low Re regimes, the required torque $\tau_{max}$ to turn a thin circular disk with radius $a$ in a background liquid with dynamic viscosity of $\mu$ can be calculated as $\tau_{max}=6\mu a^3\omega$ \cite{Happel12},
where $\omega$ is the angular velocity of the disk. Based on the above equation, in order to achieve a maximum angular speed of 25rad/s with a 4cm diameter thin disk in a liquid with 5000cP viscosity, a 6N.mm torque is needed. Considering this maximum torque along with complexity in driving\footnote{For example, brushless DC motors require a complex driver system while brushed DC motor can be driven simply by applying a voltage difference.}, cost and compactness, a geared DC motor \cite{motor} was selected which has a 6mm diameter, 19mm length and can provide 6N.mm torque\footnote{By consuming 180mA.} at 5V (see Fig. \ref{fig44}). Three dual H-bridge DC motor controllers (TB6612FNG from Toshiba) are used to control these motors in both directions. Each driver works with a 5V electric power supply, and is capable of applying a maximum continuous current up to 1.2A. Motor angular speed can be controlled through a PWM signal where the duty ratio determines the angular speed of the motor.

\subsection{Sensory System}
An accurate angular position and phase lag between the swimmer paddles and the chassis expansion is crucial to achieve propulsion and navigation in the background liquid. Consequently, an effective sensory system is necessary to measure the relative angles between each paddle and the swimmer body as well as chassis expansion length. Here, we used the AS5048A rotary magnetic encoder chip \cite{encoder}, which measures the absolute position of the magnet's rotation angle and consists of four Hall effect sensors located at its corners. It has a resolution of 14bit per revolution (i.e. 0.02 degree) and a small package size (4x6x1mm). 

\begin{figure}[h]
	\centering
	\includegraphics[width=0.45\textwidth]{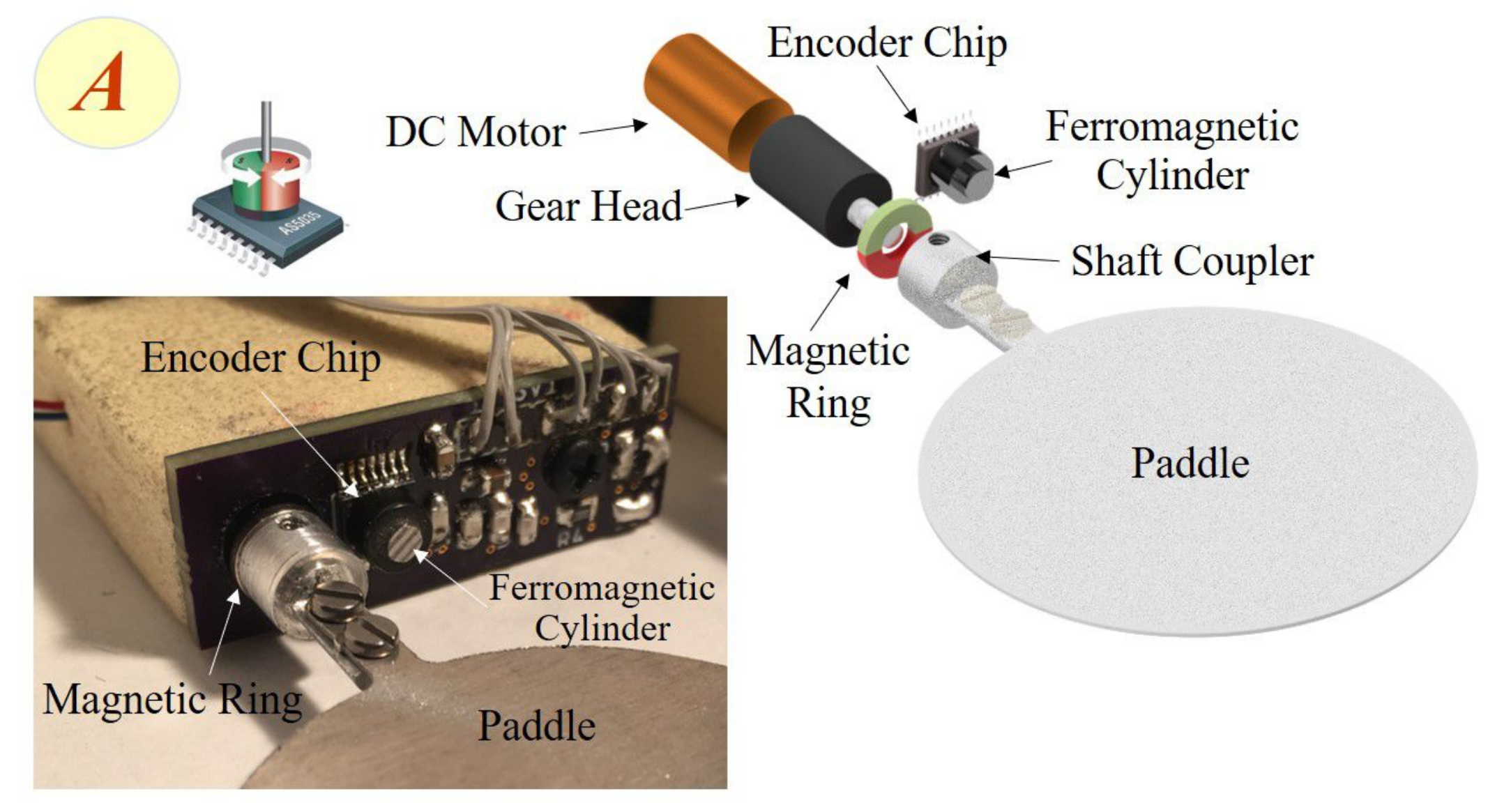} 
	\caption{A magnetic encoder is used to measure the paddle angle. The ring magnet attached to the paddle shaft induces a magnetic field in the ferromagnetic cylinder which is installed on top of the magnetic encoder chip.}
	\label{fig55}
\end{figure}

To measure the paddle angle, a magnetic ring\footnote{The ring needs to be radially magnetized.} needs to be attached to the paddle shaft while the encoder chip is in front of and centered with the ring. However, in our swimmer system, neither of the paddle shaft ends are available to attach a magnetic ring while it faces the encoder chip. Therefore, the chip has to be placed with an offset (8mm) from the ring (see Fig. \ref{fig55}), though the offset will disable the encoder from measuring the angle. To solve this issue, a small ferromagnetic cylinder is attached on the encoder chip. The magnetic ring on the paddle shaft will induce a magnetic field inside the cylinder. While the paddle shaft and magnetic ring combination is turned, the magnetic encoder chip can indirectly estimate the paddle angle by measuring the induced field in the ferromagnetic cylinder.

Similarly, a combination of a linear magnetic encoder (AS5304) and a multi-pole magnetic strip is used to measure the expansion/contraction of the chassis (see Fig. \ref{fig44}). The linear magnetic encoder is installed on the rear section of the chassis, while the magnetic strip is attached to the front. Expansion/contraction of the chassis generates a square wave pulse. By counting the number of pulses with the on-board micro-controller, the instantaneous length of the chassis can be measured. Considering the linear magnetic encoder resolution (160 pulses per pole pair interval) and the pole interval on the magnetic strip (4mm), the resolution of the chassis expansion measurement is equal to 0.025mm.

\subsection{Micro-Controller}
An on-board micro-controller is used to perform low-level control algorithms, calculate the command signals for the actuators and measure the sensor values. Considering the number of I/O ports necessary for the system as well as size, processor speed and availability, the Atmega2560 (from Atmel) is used here. It has an 8-bit processor and can run with a maximum clock frequency of 16MHz. A unique board is designed and fabricated for installing this micro-controller as well as other electronic components, such as motor drivers, voltage regulators and linear encoders. The fully assembled microswimmer system is shown in Fig. \ref{fig0}.

\section{Control Architecture}
\label{control}
\begin{figure}[h]
\begin{center}
\includegraphics[width=0.46\textwidth]{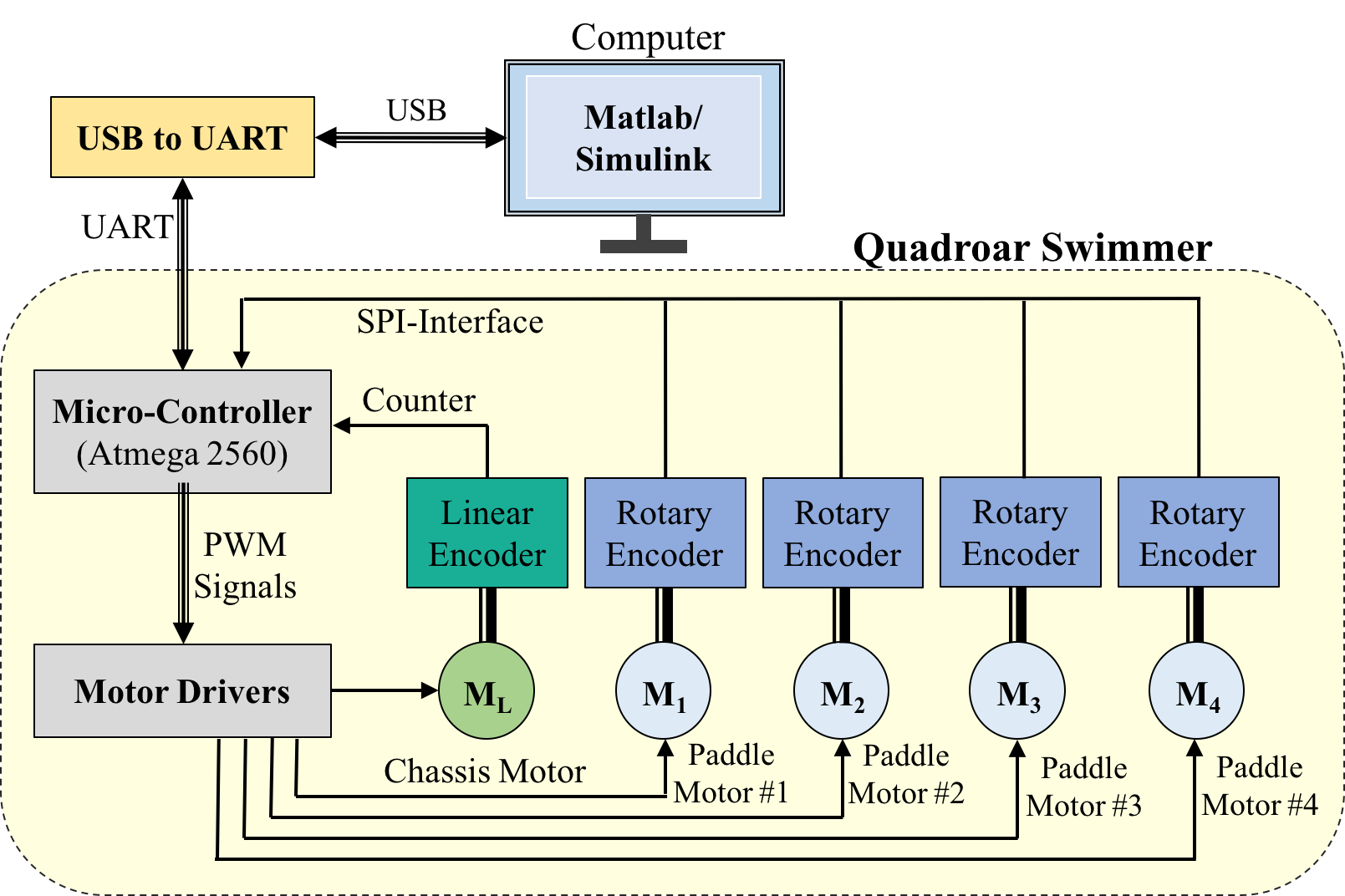}
\caption{Block diagram presenting interconnections between different components of the swimmer system.}
\label{fig5}
\end{center}
\end{figure}

The sampling rate between the micro-controller and sensors/actuators is more than 300Hz, while the communication rate between the micro-controller and the computer (attached via USB cable) is only 50Hz. Therefore, a hierarchical control scheme (c.f.  \cite{multi_rate_2}) is used here to improve the performance of the trajectory tracking (see Fig. \ref{fig5}). The two-level control structure in this paper is a multi-rate system with the higher level controller (i.e. Simulink) updating the reference signal at 50 Hz and the lower level controller (i.e. on-board micro-controller) closing the loop at 300 Hz \cite{multi_rate_1}. This control architecture improves the system performance by a fast and accurate tracking of the paddle angles and chassis expansion/contraction.

\begin{figure}[h]
	\centering
	\includegraphics[width=0.48\textwidth]{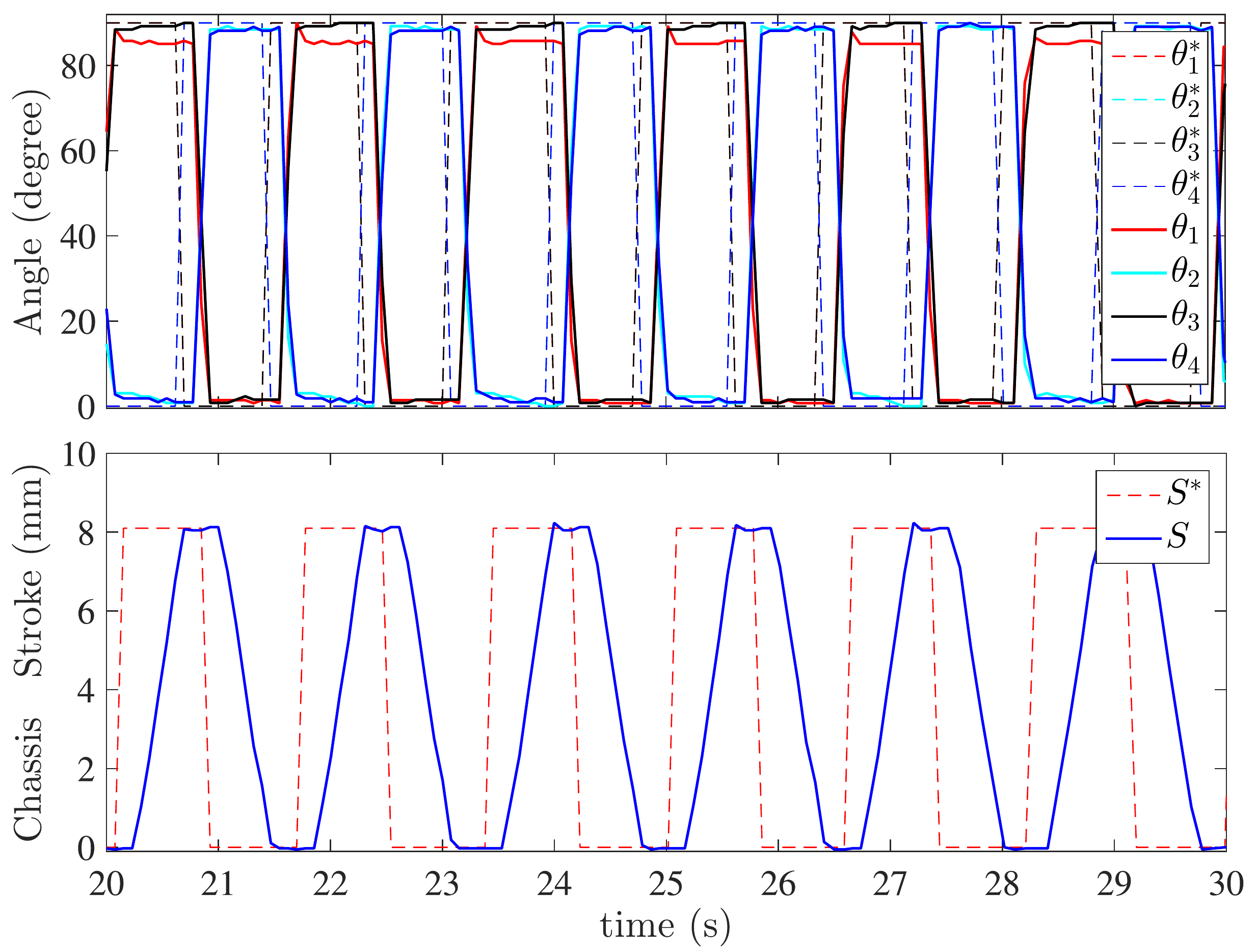} 
	\caption{Trajectory tracking for a square wave signal (both paddles and the chassis). The paddles rotate between 0 and $90^o$ while the chassis expands/retracts between 0 and 8.1mm.}
	\label{fig7}
\end{figure}

The low-level controller is composed of a feedback and a feedforward loop. The feedforward portion is obtained by an open-loop calibration test that is performed on each motor individually. The calibration is mainly done to find and remove the dead-zone in the motor control command. It is also needed to find a linear input/output model for each actuator. The feedback portion is a PID controller that has been tuned based on the motor model as well as the hydrodynamic interaction between the paddle and background liquid. The performance of this controller for trajectory tracking (here a square wave) is shown in Fig. \ref{fig7} for a sample case. This two-level control scheme is then implemented on the swimmer to test various modes of operation in silicone oil where each mode is composed of a unique sequence of paddles/chassis actuation profile. The results of these experimental tests are presented in the next section.


\section{Experimental Results}
\label{results}
\subsection{Comparison between the experiment and theory}
\begin{figure}[h]
	\centering
	\includegraphics[width=0.42\textwidth]{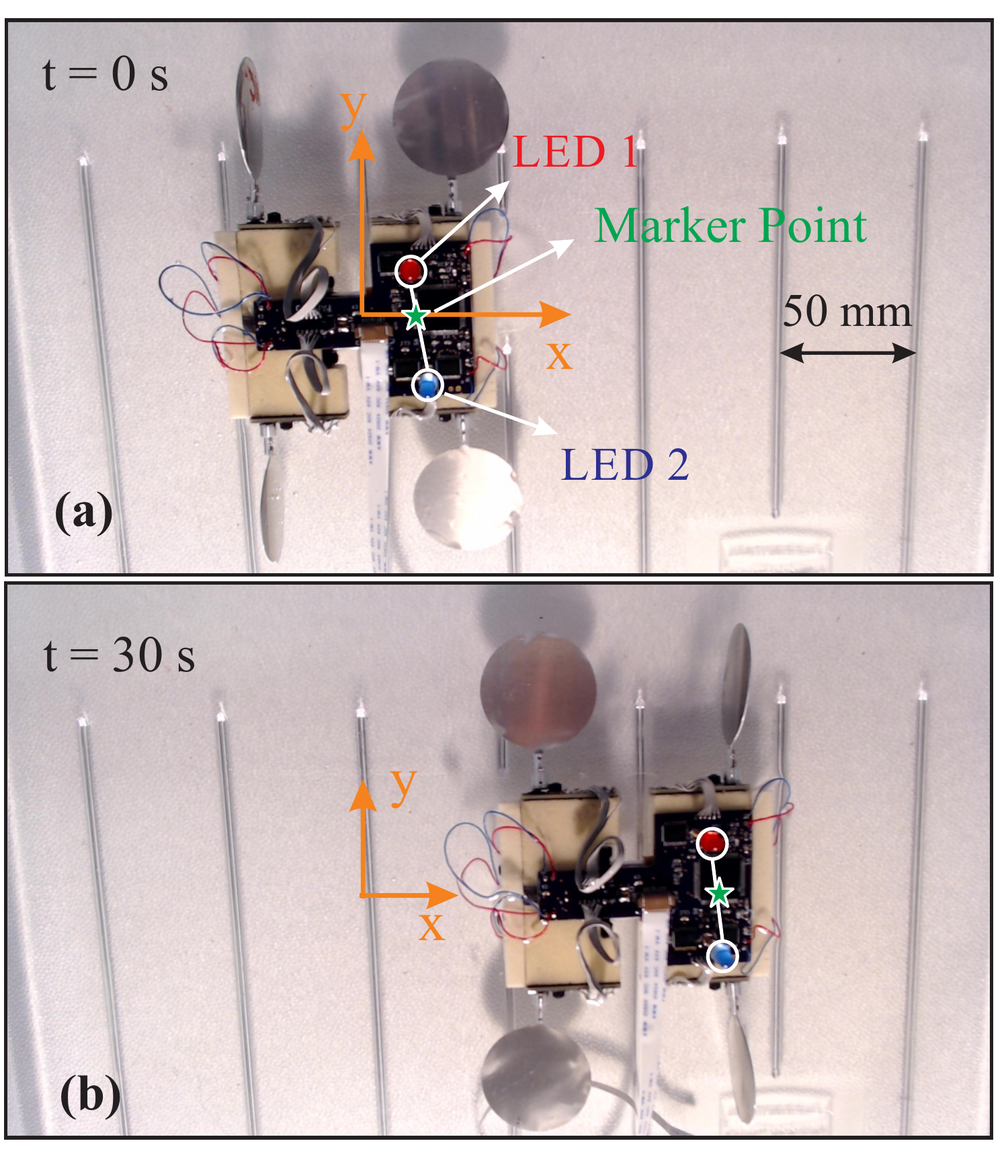} 
	\caption{The Quadroar swimmer performing forward translation mode in the silicone oil tank. Two LEDs are used to find the location and orientation of the swimmer using video processing. The stroke and linear velocity of the chassis expansion/contraction are 21.6 mm and 10.5 mm/s.}
	\label{fig6}
\end{figure}

\begin{figure}[h]
	\centering
	\hspace{10mm}\includegraphics[width=0.43\textwidth]{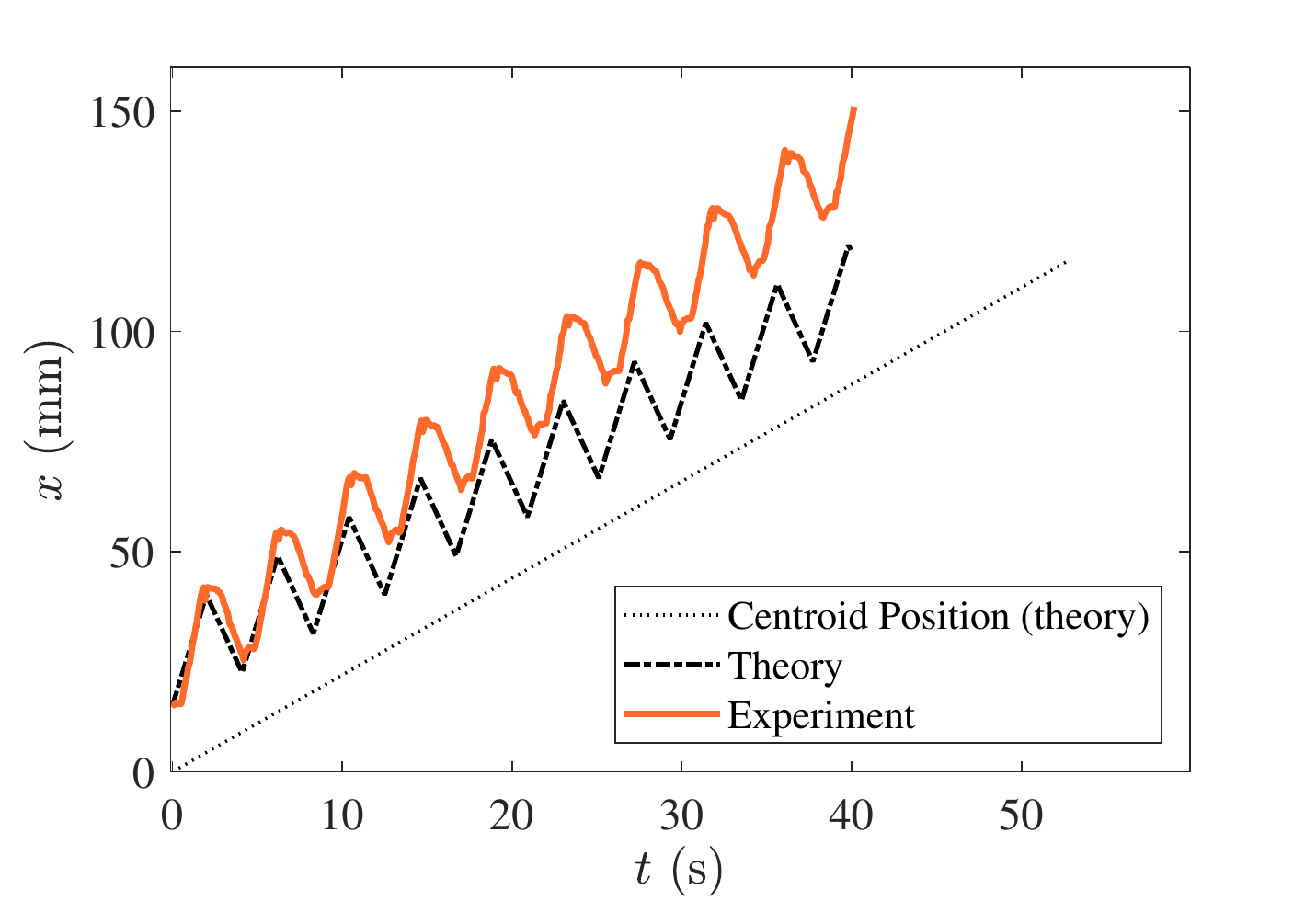} 
	\includegraphics[width=0.45\textwidth]{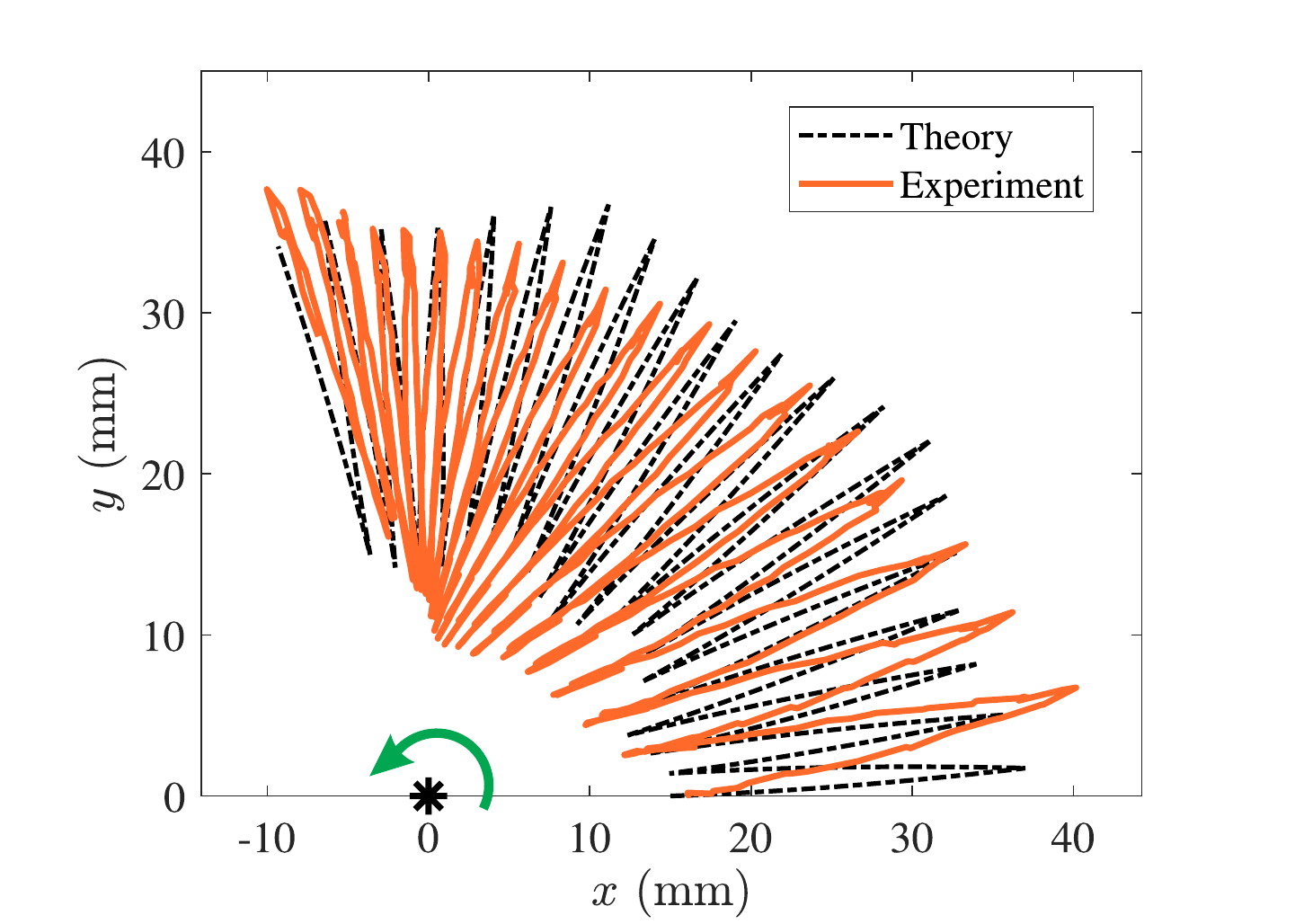} 
	\caption{Comparison between the theoretical and experimental results for sample cases in forward translation mode (top) and planar reorientation mode (bottom). Solid and dashed lines represent trace of the marker on the swimmer as obtained in experiments and theoretical modeling, respectively. Theoretical position of centroid in bottom panel is shown by an asterisk, and the green arrow shows direction of reorientation. The stroke and linear velocity of chassis' expansion/contraction are 21.6 mm and 10.5 mm/s, respectively.}
	\label{fig8}
\end{figure}

After identifying and resolving minor mechanical and electrical bugs in the device, the swimmer system has been submerged and tested in a tank of silicone oil. Experimental tests for the two primary modes (i.e. forward translation and planar reorientation) have been conducted, and the dynamic behaviour of the swimmer system is compared  with the theoretical model developed in our past studies \cite{Jalali2014}. Several videos of these experimental tests can be found at \cite{Vids}. Two time snapshots of the forward translation experiment are shown in Fig. \ref{fig6}. As shown, red and blue LEDs are used on the swimmer for determining its position and orientation in the tank (through video processing). A sequence of paddle angles and chassis stroke similar to what is shown in Fig. \ref{fig1_2} is applied on the swimmer to achieve translation and reorientation motions. The Reynolds number corresponding to these tests can be calculated as $Re=\rho_{l} v_{s} D_{p}/\mu_l$,
where $\rho_l$ and $\mu_l$ are the background liquid density and dynamic viscosity (1000Kg/m$^3$ and 5000cP), $v_s$ is the maximum chassis expansion/contraction speed (10.5 mm/s) and $D_p$ is the paddle diameter (40mm). Therefore, the Reynolds number corresponding to these experiments is $\sim 0.1$, which matches our assumption for low-Reynolds-number condition (i.e. $Re \ll 1$). The results of these two tests are compared with theoretical results in Fig. \ref{fig8}. The experimental results match well with theoretical prediction for both the translation and reorientation modes. The slight difference between the experimental and theoretical results is mainly due to neglecting the following factors in the theoretical model: (i) swimmer frame size which generates additional drag force; (ii) effect of the tank's walls and bottom surface; and (iii) tension in the data/power cable connected to the swimmer.

The match between the theoretical and experimental results opens up a new avenue to design control algorithms and obtain complex trajectory tracking in real time. This is the main topic of our future work, where we will design and implement a Passive Velocity Field Control (PVFC) algorithm to move the swimmer along any prescribed trajectory in a liquid tank with several solid obstacles to be avoided.

\subsection{Experimental exploration of different chassis stroke}

\begin{figure}[h]
	\centering
	\includegraphics[width=0.48\textwidth]{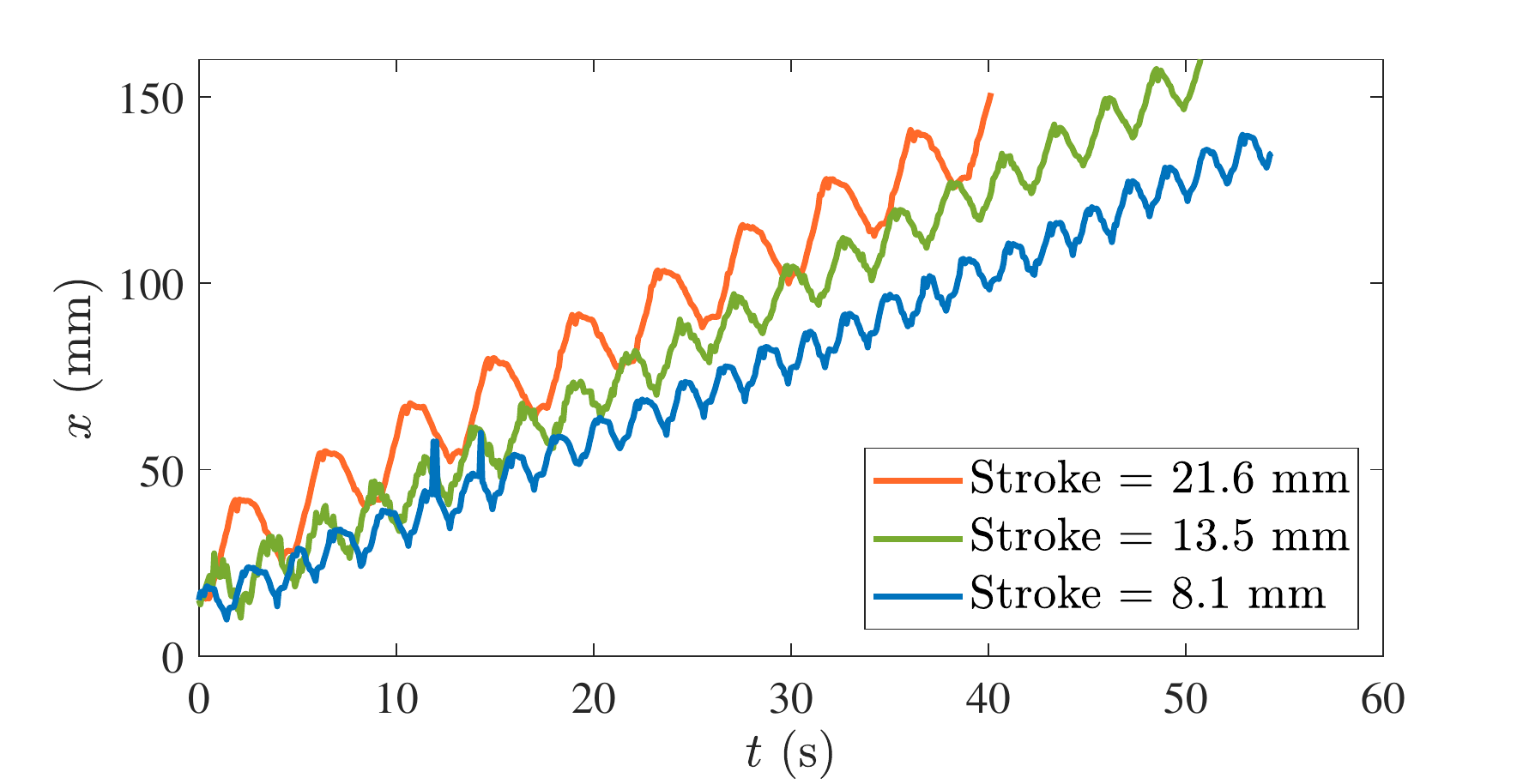} 
	\includegraphics[width=0.48\textwidth]{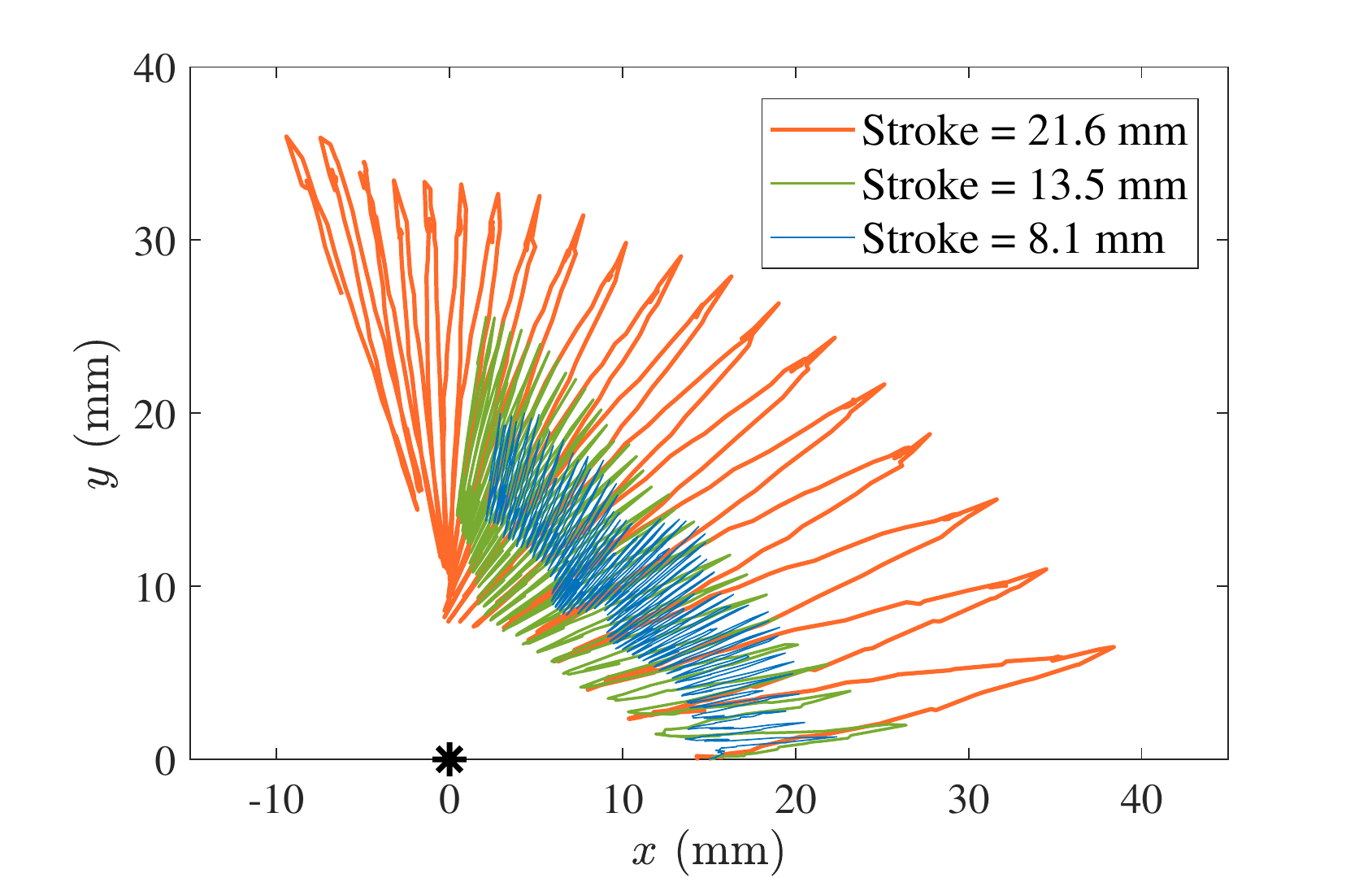} 
	\caption{Experimental results for forward translation mode (top) and planar reorientation mode (bottom) of the swimmer for different strokes. \textit{Top}: The swimmer is translating forward in x direction (c.f. Fig. \ref{fig1_2}-a); \textit{Bottom}: The swimmer is reorienting counter-clockwise in x-y plane (c.f. Fig. \ref{fig1_2}-b). Solid lines represent trace of the marker on the swimmer.}
	\label{fig10}
\end{figure}

More experimental tests have been done to investigate the effect of chassis stroke on forward translation and planar reorientation. Three different chassis strokes are used for each mode: 8.1, 13.5 and 21.6mm. As shown in Fig. \ref{fig10}, the chassis stroke does not affect the average linear velocity of the swimmer. This is in agreement with our theoretical prediction, which shows that the swimmer linear speed only depends on the chassis expansion/contraction rate (i.e. speed) but not its stroke. In planar reorientation, note that trajectories belong to the marker point (Fig. \ref{fig6}), since motion of the swimmer's centroid is negligible in our experiments.


\section{CONCLUSIONS}
\label{conclude}

We designed, fabricated, and experimentally tested a mm-scale low-Reynolds-number swimmer, called the Quadroar. The swimmer was first introduced in our previous works, and is composed of four independent paddles (Quadro+oar) and a chassis capable of expanding/retracting. The swimmer body is made with a machinable foam to make it neutrally buoyant in silicone oil. A two-level supervisory control scheme was designed and implemented on the swimmer to perform two primary modes of swimming (i.e., forward translation and planar reorientation). The supervisory controller (i.e., computer/Simulink) calculates the sequence of paddles angle as well as chassis expansion/contraction to obtain the desired motion, while the on-board controller (i.e. micro-controller) implements feedback algorithms to track those angles and stroke by the paddles and chassis, respectively. The performance of the mm-scale Quadroar swimmer has been experimentally evaluated in silicone oil, where it satisfies the low-Reynolds-number condition (Re $\sim 0.1$). The two primary modes of operation - forward translation and planar reorientation - have been tested and compared with our theoretical models. Experimental results matched well with our theoretical predictions, while some deviations observed were due to factors such as body drag force, wall effects and tension in the data/power cable, which were not considered in the theory. Our future works mainly involve the design and implementation of a Passive Velocity Field Control (PVFC) algorithm to move the swimmer robot along a predefined trajectory such that it avoids solid obstacles on its motion path.

{}


\begin{thebibliography}{}


\bibitem{Purcell77}
E.M. Purcell, ``Life at low Reynolds number" \emph{American Journal of Physics}, 45, 3, 1977.

\bibitem{Happel12}
J. Happel, and H. Brenner, \emph{Low Reynolds number hydrodynamics: with special applications to particulate media}, Vol. 1, Springer Science \& Business Media (2012)


\bibitem{Najafi04}
A. Najafi, and R. Golestanian, ``Simple swimmer at low Reynolds number: Three linked spheres" \emph{Physical Review E}, 69, p. 062901, 2004.

\bibitem{Zhang09}
L. Zhang, J.J. Abbott, L. Dong, B.E. Kratochvil, D. Bell, and B.J. Nelson, ``Artificial bacterial flagella: Fabrication and magnetic control" \emph{Applied Physics Letters}, 94, p. 064107, 2009.

\bibitem{Jalali2014}
M.A. Jalali, M.R. Alam, and S. Mousavi, ``Versatile low-Reynolds-number swimmer with three-dimensional maneuverability" \emph{Physical Review E}, 90(5), p. 053006, 2014.

\bibitem{Tottori12}
S. Tottori et al., ``Magnetic helical micromachines: fabrication, controlled swimming, and cargo transport" \emph{Advanced materials}, 24, 2012.

\bibitem{Mirzakhanloo2019}
M. Mirzakhanloo, and M.-R. Alam, ``Concealed Swarm of Micro-swimmers", \emph{Preprint}, arXiv:1811.10101, 2018.

\bibitem{Jalali2015}
M.A. Jalali, A. Khoshnood, and M.-R. Alam, ``Microswimmer-induced chaotic mixing" \emph{Journal of Fluid Mechanics}, 779, 669, 2015.

\bibitem{Katija2012}
K. Katija, ``Biogenic inputs to ocean mixing" \emph{The Journal of experimental biology}, 215, pp. 1040-1049, 2012.

\bibitem{Wilhelmus2014}
M. M. Wilhelmus, and J. O. Dabiri, ``Observations of large-scale fluid transport by laser-guided plankton aggregations" \emph{Physics of Fluids}, 26, p. 101302, 2014.

\bibitem{Mirzakhanloo2018A}
M. Mirzakhanloo, M.A. Jalali, and M.-R. Alam, ``Hydrodynamic Choreographies of Microswimmers" \emph{Scientific reports}, 8(1), 3670, 2018.

\bibitem{Mirzakhanloo2018B}
M. Mirzakhanloo, and M.-R. Alam, ``Flow characteristics of Chlamydomonas result in purely hydrodynamic scattering" \emph{Physical Review E}, 98(1), p. 012603, 2018.

\bibitem{Medina2017}
M. Medina-S\'anchez and O. G. Schmidt, ``Medical microbots need better imaging and control" \emph{Nature}, 545, pp. 406-408, 2017.

\bibitem{multi_rate_2}
Barcelliy, D., Bemporadz, A. and Ripaccioliy, G., ``Hierarchical multi-rate control design for constrained linear systems," \emph{49th IEEE Conference on Decision and Control (CDC)}, pp. 5216-5221, 2010.

\bibitem{multi_rate_1}
Hong, H.J., Li, Z.Q., Fan, S.X., Zhou, Q.K. and Fan, D.P., ``Multi-rate tracking controller analysis and design for target tracking systems," \emph{Springer}, 20(11), pp.3049-3056, 2013.

\bibitem{motor}
Brushed DC Gear Motor, Available: \url{https://www.precisionmicrodrives.com/
product/206-109-6mm-dc-gearmotor-19mm-type} [Accessed: June, 2017].

\bibitem{encoder}
Austria Micro Systems AS5050A Rotary Position Sensor,
Available: \url{http://ams.com/eng/Products/Magnetic-Position-Sensors/
Angle-Position-On-Axis/AS5050A} [Accessed: June, 2017].

\bibitem{Vids}
See videos at: \url{https://sites.google.com/berkeley.edu/mohsen-saadat/projects/low-reynolds-number-swimmer?authuser=1}

\end{thebibliography}
\end{document}